\documentclass[%
 amsmath,amssymb,
 aps,
 pre,
 twocolumn,
 superscriptaddress,
 showpacs,
 showkeys
]{revtex4-1}

\usepackage{graphicx}
\usepackage{dcolumn}
\usepackage{bm}
\usepackage{xcolor}
\usepackage{color}
\usepackage{mathrsfs}

\begin{document}

\title{Coulomb force mediated heat transfer in the near field - geometric effect} 

\author{Jian-Sheng Wang}
\affiliation{Department of Physics, National University of Singapore, Singapore 117551, Republic of Singapore}
\author{Zu-Quan Zhang}
\author{Jing-Tao L\"u}
\affiliation{School of Physics and Wuhan National High Magnetic Field Center, Huazhong University of Science and Technology, 430074 Wuhan, P. R. China}

\date{5 March 2018}

\begin{abstract}
It has been shown recently that the Coulomb part of electromagnetic interactions
is more important than transverse propagation waves for the near-field enhancement of 
heat transfer between metal objects  at a distance of order nanometers.   
Here we present a theory focusing solely on the Coulomb
potential between electrons hopping among tight-binding sites.   When the relevant systems are
reduced to very small geometry, for example, a single site, the enhancement is much higher
compared to a collection of them packed within a distance of a few angstroms.   We credit this to the
screening effect.   This result may be useful in designing metal-based meta-materials to enhance
heat transfer much higher.
\end{abstract}
\maketitle

\section{Introduction}
During the 1970s, it was discovered both experimentally \cite{hargreaves69,domoto70} 
and explained theoretically \cite{PvH} that
the transfer of thermal radiative energy between two plates is enhanced greatly when the distances
are of the order of the thermal de Broglie wave length, which near room temperature is about
several micrometers.   The theoretical prediction is based on Maxwell's equations coupled to
the random thermal motion of the sources, which obeys
the fluctuation-dissipation theorem of Callen and Welton \cite{callenwelton51} 
when a subsystem is in thermal equilibrium.   Using these fundamental ideas
with two subsystems each at a different temperature, a heat flux is predicted.   

The heat current has now been observed down to much smaller distance scales.
Recent experiments have 
approached distances of the order of one nanometer or smaller
\cite{kittle05,kim2015radiative,song2016radiative,Cui_NatCom2017,Kloppstech2015}.   At these short distances, the Coulomb
interaction represented by the scalar potential is more important than the propagating field represented
by the vector potential (say in a Coulomb or transverse gauge) \cite{Keller11}.   Although the $\bf E$ and $\bf B$ field 
themselves are gauge independent in Maxwell's 
equations, it is economical to consider only
the instantaneous Coulomb charge interactions and ignore the retardation part of the field.
Indeed, this has been done in a number of papers \cite{Yu2017,Mahan2017,Wang2017,Jiang2017,jiebin2017,zuquanPRB}.   The usual approach along the line of
Polder and van Hove (PvH) \cite{PvH} or its generalization \cite{volokitin07,Basu09,Song-rev15} is to separate 
the question into two problems, the first one is a material
property problem, where the dielectric function is determined, the second is to solve the
Maxwell equations.   However, we find it more appropriate if the problem is formulated from
the start as a condensed matter physics problem with a given Hamiltonian.   This allows geometric
consideration to  be put in naturally without relying on other factors, for example, the locality approximation for
the dielectric function.  In fact, at these very short distances, the long-wave limit result for the
dielectric function is not expected to be valid. 

In this paper, we will give a brief outline of a theory based on nonequilibrium Green's function (NEGF) \cite{haug96,wang08review,wang14rev}
to compute the energy
transport for the Coulomb systems.  This is based on the Meir-Wingreen formula for total energy 
current, which can be shown to be reducible to a Landauer-like expression with a Caroli formula as the
transmission coefficient.   In appendix A, we give an alternative derivation based on fluctuational electrodynamics,
and in appendix B, we show that it is identical to that of Yu et al.~\cite{Yu2017}.   We apply the formalism to 
three cases, two quantum dots with three-dimensional Coulomb interaction, a quantum dot with a surface
of a cubic lattice, and two cubic lattices with varying cross-sectional areas.   We also discuss the
layer number dependence when the central region is enlarged.    The main conclusion of
the work is that geometry at the atomic scale plays a major role in giving a very large 
heat transfer.   This is mainly due to the fact that when systems are small, screening is not
effective, thus unscreened point charges carry large energy currents.

\section{Coulomb interaction electron model} \label{THEORY} 
We consider the following Hamiltonian for a collection of point charges
in three dimensional space interacting through the Coulomb potential,
\begin{equation}
\label{eq-Hamiltonian}
\hat{H} = c^\dagger H c + \frac{e^2}{2} \sum_{i,j} c_i^\dagger c_j^\dagger v_{ij}
c_j c_i.
\end{equation}
Here $c$ is a column vector where each entry is the annihilation operator $c_j$ on a discrete site $j$, while
$c^\dagger$ row vector of their Hermitian conjugates.
$H$ is  a Hermitian matrix, $H = H^\dagger$, which will  be separated as system or center,
$H_C$, and any number of electron baths, $H_B^\alpha$, and  their couplings, $V_{CB}^\alpha$, as 
submatrices.   For simplicity of NEGF treatment, and also well justified by the screening
property of Coulomb interaction, we assume the Coulomb interaction $v_{ij}$ occurs only
for the sites within the center region.  Thus, the electron baths or leads will be ``free'' electrons.
As far as the formal theory goes, the interaction matrix $v_{ij} = v_{ji}$ is a real symmetric matrix with 
arbitrary values.   Note that if $i=j$, since $(c_j)^2 = 0$, the diagonal terms are never needed.  So for
convenience, we define $v_{ii} = 0$.  The self-interaction is forbidden due to Pauli
exclusion principle.  Note also that our model of the electrons has no spins.  In three dimensions,
for point charges, we take
\begin{equation}
(D^0)_{ij} = v_{ij} = \frac{1}{ 4 \pi \epsilon_0 r_{ij}},
\end{equation}
where $\epsilon_0$ is the dielectric constant of vacuum, and $r_{ij}$ is the Euclidean distance
between site $i$ and $j$.  Equation (\ref{eq-Hamiltonian}) is standard and forms the starting point
of many theoretical developments, such as in Kadanoff and Baym \cite{kadanoffbaym}, or in Mahan \cite{mahan00}. 

\section{NEGF method for energy currents}
To study heat transport, we need to solve a Dyson equation for the scalar field Green's function,
$D = D^0 + D^0 \Pi D$, or more precisely, this is defined on the Keldysh contour
with space index $j$ and contour time $\tau$, i.e.,
\begin{eqnarray}
D_{ij}(\tau, \tau') &=&  
D^{0}_{ij}(\tau, \tau') +  \\ 
&&\!\!\!\!\!\! \sum_{k,l} \int\!\!\int d\tau_1 d\tau_2 D^{0}_{ik}(\tau, \tau_1) \Pi_{kl}(\tau_1, \tau_2)
D_{lj}(\tau_2, \tau').\nonumber 
\end{eqnarray}
Here the contour function $D$ is the screened Coulomb potential, and is usually denoted as $W$ in many-body
theory, and $\Pi$ is the polarization function or scalar photon self-energy.    Since the Coulomb
interaction is instantaneous in real time, we must have $D^0(\tau, \tau') \propto \delta(\tau, \tau')$.
As a result, we do not have lesser or greater components, $(D^0)^< = (D^0)^> = 0$, and
$(D^0)^r = v\, \delta(t-t')$.   The contour Dyson equation is then reduced to a retarded one,
$D^r = v + v \Pi^r D^r$, and the Keldysh equation, $D^< = D^r \Pi^< D^a$, which is most
conveniently handled in the angular frequency domain, for example, 
\begin{equation}
D^r(\omega) = \int_{-\infty}^{+\infty}\!\!\!\! D^r(t) e^{i \omega t} dt.
\end{equation}

\subsection{Energy current formulas}
We consider a two-terminal or two-bath situation labelled as 1 and 2, and assume that electrons cannot
jump from one side to the other, i.e., the Hamiltonian $H$ is block diagonal.  We can  then derive
a Caroli formula \cite{Caroli1971}, of the energy current out of the lead 1,
\begin{eqnarray}
\label{eq-landauer}
I_1 &=& \int_0^{+\infty}\!\! \frac{d \omega}{2 \pi} \hbar \omega\, T(\omega) \bigl( N_1 - N_2\bigr),\\
\label{eq-caroli}
T(\omega) &=& {\rm Tr} \bigl( D^r \Gamma_1 D^a \Gamma_2  \bigr).
\end{eqnarray}
Here $N_\alpha = 1/\bigl[ \exp(\beta_\alpha \hbar \omega) - 1\bigr]$ is the Bose function at the temperature 
$T_\alpha = 1/(k_B \beta_\alpha)$
of the lead $\alpha$, $k_B$ is the Boltzmann constant, the spectrum function is defined as
$\Gamma_\alpha = i (\Pi_\alpha^r - \Pi_\alpha^a)$, $\alpha=1,2$.   Since there is no explicit coupling of the electrons
at least at the random phase approximation (RPA) level,  $(\Gamma_1)_{ij}$ is 0 unless both space indices $i,j$  are on the same side indexed by 1.
Thus the above procedure gives a quick recipe to compute the heat current.   It was shown in Ref.~\onlinecite{zuquanPRB} that
this  Caroli formula agrees with the usual fluctuational electrodynamics in the non-retardation limit, and 
is derivable approximately from a more rigorous Meir-Wingreen formula.

The Caroli formula is not valid when the two sides are coupled electronically and electrons interact by
Coulomb interaction.   For such situations, we need to use the more general Meir-Wingreen formula \cite{meir-wingreen92}
given by
\begin{equation}
\label{eq-meir-wingreen}
J_\alpha = \int_{-\infty}^{+\infty} \frac{dE}{2\pi \hbar} 
E\, {\rm Tr} \Bigl(G^> \Sigma_\alpha^< - G^< \Sigma_\alpha^>   \Bigr).
\end{equation}
Here, the trace is over the space indices, and the electron Green's functions and lead $\alpha$ self-energies
are functions of energy $E = \hbar \omega$.   The electron Green's function satisfies
a similar Dyson equation as for $D$.  However, the electron self-energy $\Sigma$ 
with the Coulomb interaction in action
cannot be obtained exactly.   Various approximate schemes are available, such as the Hartree-Fock method,
self-consistent Born approximation \cite{lu2007} (or more commonly known as GW method), and 
the formal Hedin equations \cite{hedin65}.   To show the equivalence
of Eq.(\ref{eq-meir-wingreen}) with (\ref{eq-landauer}) and (\ref{eq-caroli}), we need the following
conditions:
(1)  Electrons cannot move from one side to other, and thus we assume the electron Green's functions
$G$ as well as photon self-energy $\Pi$ are block diagonal.  (2) In applying the Keldysh equation,
we take the lowest order approximation for the retarded/advanced Green's functions, i.e.,
we use $G^< \approx G_0^r  \Sigma_{\rm tot}^< G_0^a$.  Here subscript 0 means the Coulomb
interaction is turned off for the electrons.  With the assumptions (1) and (2), we can show $J_1 = I_1$
exactly. 

\subsection{Random phase approximation for photon self-energy}

For the rest of the texts we will focus on the application of the Caroli formula under the assumption that the electrons
are not directly coupled.   The materials property is then uniquely defined through the retarded
scalar photon self-energy $\Pi^r$.  This quantity is easily expressed in time domain as
$\Pi^r(t) = \theta(t) \bigl( \Pi^>(t) - \Pi^<(t) \bigr)$, with the matrix elements
\begin{equation}
\Pi_{jk}^>(t) = - i \hbar  e^2 G_{0,jk}^>(t) G_{0,kj}^<(-t).
\end{equation}
$\Pi^<$ is obtained by a swap $>\, \leftrightarrow \, <$.   The above expression represents the
lowest order Dyson expansion for the self-energies, and is known as RPA \cite{mahan00,bruus04}.   The electron
Green's functions, $G_0^{>,<}$, are evaluated when the Coulomb interaction is absent. 
The time domain expression is convenient for fast Fourier transform.  However, it is
not necessarily more efficient or more accurate, since the spacings or range in time or frequency 
cannot be chosen at will.   Alternatively we can compute directly in the frequency domain
with the formula
\begin{eqnarray}
\Pi^r_{jk}(\omega) &=& - i \hbar e^2 \int_{-\infty}^{+\infty}
\frac{dE}{2\pi \hbar} \Big[ 
G^r_{0,jk}(E) G^<_{0,kj}(E-\hbar \omega) + \nonumber \\
\label{eq-RPA}
&& \qquad G^<_{0,jk}(E) G^a_{0,kj}(E-\hbar \omega) \Big]. 
\end{eqnarray} 
The lesser component of the electron Green's function is then 
calculated with the fluctuation-dissipation relation, 
$G_0^< = - f (G_0^r - G_0^a)$, where $f = 1/\bigl[ \exp\bigl(\beta_\alpha(E -\mu_\alpha)\bigr) +1 \bigr]$
for the side connected to the $\alpha$-th lead (remember $G_0$ is block diagonal). 
The retarded Green's function is obtained by solving the Dyson equation with the surface Green's functions,
$G_0^r(E) = (E - H_c - \sum_\alpha \Sigma_\alpha)^{-1}$. 

\begin{figure} 
\includegraphics[width=\columnwidth]{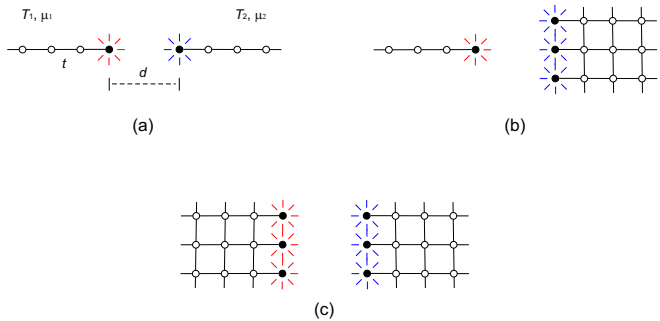}
\caption{Illustration of the models. (a) 1D lattices with Coulomb interaction in three dimensions.
(b) 1D chain with a surface  of cubic lattice (with $L=3$ in the drawing).  (c) Two surfaces of cubic lattices.
Solid dots indicate sites that have Coulomb interactions, while open circles belong to the leads.
For simplicity of the drawings, only a cross-section is illustrated in (b) and (c).}
\label{fig-illustrate}
\end{figure} 
\section{Applications}
We present some applications of the above formalism to simple systems.  Fig.~\ref{fig-illustrate}
gives a schematic illustration of the models. 

\subsection{Two quantum dots in three dimensions}
We consider a 1D chain terminating at $(0,0,0)$ on the left, and another 1D chain starting at
$(0,0,d)$, with lattice constant $a$ and hopping parameter $t$.  Only the end points experience 
Coulomb interaction and the rest of the chains
serve as leads.   Since the point charges are treated as in three dimensions, the 
Coulomb interaction matrix is
\begin{equation}
v = \left( \begin{array}{cc}
0 &  \frac{1}{4\pi \epsilon_0 d} \\
\frac{1}{4\pi \epsilon_0 d}    & 0
\end{array}   
\right).
\end{equation}
Since we have only two sites  in the center system (call  them 1 and 2), the $v$ matrix is 2 by 2.
As a side comment, note that this differs from a capacitor for which we have 
\begin{equation}
v^{-1} = \left( \begin{array}{cc}
C & -C \\
-C  & C
\end{array}   
\right), \quad C = \frac{\epsilon_0 A}{d}.
\end{equation}
Here $C$ is the capacitance of a parallel plate capacitor distance $d$ apart and on area $A$. 
Since $v^{-1} \phi = q$, where $q = (q_1, q_2)^T$, and $\phi = (\phi_1, \phi_2)^T$, 
this represents the simple physics of a capacitor, where  $q_1 = - q_2 = (\phi_1 - \phi_2)C$.
Although for the capacitor, $v^{-1}$ is not invertible, we can represent it as
\begin{equation}
v = \frac{1}{2 \epsilon_0 A \eta} \left( \begin{array}{cc}
1 &  e^{-\eta d}  \\
e^{-\eta d}   & 1
\end{array}   
\right), \quad \eta \to 0^{+}.
\end{equation}

For the two-dot model in 3D, the Dyson equation,
$D^r = v + v \Pi^r D^r$ in component form is
\begin{eqnarray}
D_{11} = v \Pi_{2} D_{21}, &\quad&
D_{12} = v + v \Pi_{2} D_{22}, \\ 
D_{21} = v + v \Pi_{1} D_{11}, &\quad& 
D_{22} = v \Pi_{1} D_{12}.
\end{eqnarray}
In the above and below, for notational simplicity, we use 
$v=1/(4\pi \epsilon_0 d)$ to denote a scalar instead of the Coulomb matrix.
$\Pi^r$ is diagonal,
\begin{equation}
\Pi^r = \left( \begin{array}{cc}
\Pi_1 &  0 \\
0   & \Pi_2
\end{array}   
\right),
\end{equation}
and for matrix elements, we have dropped the superscript $r$.   
The solutions are easily obtained, such as  $D_{12} = D_{21} = 
v/(1 -v \Pi_1 v \Pi_2)$.
We can compute the transmission function from the Caroli formula as
\begin{equation}
T(\omega) = \frac{ 4 v^2 {\rm Im} \Pi_1 {\rm Im} \Pi_2 }{\big|  1 - v^2 \Pi_1 \Pi_2  \big|^2}.
\end{equation}

Although we do not have explicit results for the self-energies $\Pi_\alpha$, the following approximation
gives very accurate results,
\begin{equation}
\Pi_1 \approx \Pi_2 \approx  - \frac{e^2}{\Gamma} - i \delta \frac{e^2}{\Gamma^2} \hbar \omega.
\end{equation}
Here $\Gamma$ supplies an energy scale of order eV, but the scalar photon energies contributing
to the energy transport is  of
the order of $k_B T$.  At room temperature, this is a much smaller number, thus we can take
small $\omega$ expansion and leave only up to the linear term.   
Fitting the 1D chain result with hopping parameter $t=0.85\,$eV near room temperature gives
$\Gamma \approx 2.025\,$eV and $\delta \approx 1.750$.
We also drop the imaginary
part contribution in the denominator in the transmission formula and use the approximation,
\begin{equation}
T(\omega) \approx \frac{ 4  v^2 \left(\delta e^2\hbar \omega/\Gamma^2\right)^2}{\left(1 - v^2 e^4/\Gamma^2\right)^2}. 
\end{equation}  
With this approximation to the transmission, the Landauer formula can be integrated, which 
has the same form as the Planck blackbody radiation formula.   We obtain
\begin{eqnarray}
I_1 &=&  \int_0^{+\infty}\!\! \frac{d \omega}{2 \pi} \hbar \omega\, 
\frac{ 4  v^2 \left(\delta e^2\hbar \omega/\Gamma^2\right)^2}{\bigl(1 - v^2 e^4/\Gamma^2\bigr)^2}
\bigl( N_1 - N_2\bigr)  \nonumber\\
&=& \left( \frac{1}{\beta_1^4} - \frac{1}{\beta_2^4} \right) 
\frac{ 4 \left(\delta v e^2/\Gamma^2\right)^2}{\bigl(1 - v^2 e^4/\Gamma^2\bigr)^2}
\frac{1}{2\pi \hbar} \int_0^\infty\!\! dx \frac{x^3}{e^x-1} \nonumber \\
&=& \frac{ 8\pi  \left( \frac{\delta \alpha \lambda^2}{d} \right)^2}{\left[1- \left( \frac{\alpha \lambda}{d} \right)^2\right]^2} \, j_{BB}.
\label{eq-analytic}
\end{eqnarray} 
The integral has a well-known value of $\pi^4/15$, and the final result is rewritten 
in terms of the black-body result of energy flux $j_{BB} = \sigma (T_1^4 - T_2^4)$, here
the Stefan-Boltzmann constant is $\sigma = \pi^2k_B^4/(60 \hbar^3 c^2)$ and $c$ is the speed of light,
and $\alpha  = e^2/(4\pi \epsilon_0 \hbar c) \approx 1/137$  is the fine structure constant.  $\delta
\approx 2$ is dimensionless, and we have defined $\lambda \equiv \hbar c/\Gamma$
which has a dimension of length. 

The prefactor of $j_{BB}$ has the units of area.  We can interpret the result as each dot supplying
an amount of energy flux equivalent to the blackbody one of area order $\lambda^4/d^2$.  Since
$\Gamma$ is of order eV, $\lambda$ is of the order 100 nm.  The parameter $\lambda$ enters into the 
formula as a 4-th power, thus the effective enhancement is rather large.    In Fig.~\ref{fig-tip-surface} we compare the 
analytic result with numerical calculation.  We obtained an excellent agreement except at the singular point
$d \approx 0.7\,$nm.  This is due to our neglection of the imaginary part in $\Pi$ for the denominator
of the transmission function.

\subsection{Tip with a surface}
This will be a slight generalization of the two-point charge model presented above.   On the left, we still have
a 1D chain ending with the last point at the origin $(0,0,0)$ experiencing Coulomb interaction with
a surface of a cubic lattice located at $z \ge d$.   The cubic lattice is $L \times L \times
\infty$ occupying the $z$ coordinates at $d$, $d+a$, $d+2a$, etc.   Only nearest neighbor
hoppings are allowed both for the 1D chain and cubic lattice with the same lattice constant $a$ and
hopping parameter $t$,  but hoppings between the two sides are forbidden.  We choose an odd integer
for $L$ so that the point charge is centered.  Only the sites on the first layer of the cubic lattice have Coulomb interactions,
and the rest of layers are considered as a free electron bath.   We use periodic boundary conditions for
the cubic lattice in the transverse ($x$ and $y$) directions.  This also applies to the Coulomb terms.
We define a Fourier transform
\begin{equation}
d({\bf q}) = \sum_{j=2}^{L^2+1} D_{j1} e^{-i{\bf q} \cdot {\bf r}_j },
\end{equation}
where we label the left quantum dot as site 1, and right surface on the square lattice as 2 to $L^2+1$.
${\bf r}_j$ is the position vector of the site $j$ on the surface.   ${\bf q}$ is a two-dimensional
wave vector taking the discrete values within the first 2D Brillouin zone $\bigl(2\pi l_x/(aL), 2\pi l_y/(aL)\bigr)$, 
$l_x$ and $l_y$ taking integers.

With the ${\bf q}$-space Fourier transform, the Caroli formula can be written as
\begin{equation}
T(\omega) = \frac{\Gamma_{11}}{L^2} \sum_{{\bf q}}
\Gamma_{22}({\bf q}) \bigl| d({\bf q}) \bigr|^2.
\end{equation}
Here $\Gamma_{22}(\bf q)$ is the Fourier transform of real space $\Gamma_{jk}$, where
$j$ and $k$ run over the sites on the surface of the cubic lattice.  
$\Gamma_{11} = i (\Pi_{11} - \Pi_{11}^{*})$, and similarly, 
$\Gamma_{22} = i (\Pi - \Pi^{\dagger})$, where $\Pi$ is the self-energy for the right
cubic lattice surface.  All these quantities as well as $d$ are functions of frequency
$\omega$ which we have suppressed for notational simplicity. 

The Dyson equation takes the same form, $D^r = v + v \Pi^r D^r$, and now, $\Pi^r$ is block diagonal
with submatrices $\Pi_{11}$ of $1\times 1$ and $\Pi$ of $L^2 \times L^2$.  Separating out the terms with indices 
of 1 and greater than 1 
(for the left site and the right sites), noting $v_{11} = 0$, we get ($j>1$)
\begin{eqnarray}
D_{11} &=& \sum_{k,l>1} v_{1k} \Pi_{kl} D_{l1},\\
D_{j1} &=& v_{j1} + v_{j1} \Pi_{11} D_{11} + \sum_{k,l>1} v_{jk} \Pi_{kl} D_{l1}.
\end{eqnarray}
Eliminating $D_{11}$, the two equations are easily combined into one.
The surface property represented by $\Pi_{kl}$ is space translationally invariant (at least
under RPA), thus we can represent the matrix more economically by its
Fourier transform as $\Pi({\bf q})$.    We have
\begin{eqnarray}
d({\bf q}) &=& v({\bf q}) + v_{0}({\bf q}) \Pi({\bf q}) d({\bf q}) + \nonumber \\ 
\label{eq-dyson-q}
&&\qquad \Pi_{11} \frac{v({\bf q})}{L^2} \sum_{\bf p} v({-\bf p}) \Pi({\bf p}) d({\bf p}).
\end{eqnarray}
Here we have defined the Fourier transform of the Coulomb interaction between the point charge and
the surface sites $j$ as 
\begin{equation}
v({\bf q}) = \sum_{j=2}^{L^2+1} \frac{e^{-i (q_x x_j+ q_y y_j)} }{ 4 \pi \epsilon_0 \sqrt{ x_j^2 + y_j^2 + d^2}},
\end{equation}
and $v_{0}({\bf q})$ is similarly defined when $d=0$, i.e., $v_0$ is the intra-layer Coulomb interaction. 
Let $y$ be the term of the ${\bf p}$ summation in 
Eq.~(\ref{eq-dyson-q}), $y = {(1/L^2)} \sum_{\bf p} v({-\bf p}) \Pi({\bf p}) d({\bf p})$,
then the linear equations can be solved, if we put $d(\bf q)$ back into $y$.    We get
\begin{eqnarray}
z &\equiv & \frac{1}{L^2} \sum_{\bf p} \frac{ \left| v({\bf p}) \right|^2 \Pi({\bf p}) }{1 - v_0({\bf p}) \Pi({\bf p})},\\
y &=& \frac{z}{1 - z \Pi_{11}},\\
d({\bf q}) &=& \frac{ v({\bf q}) ( 1 + y\Pi_{11})}{1 - v_{0}({\bf q}) \Pi({\bf q}) }.
\end{eqnarray}
The photon self-energies are calculated from Eq.(\ref{eq-RPA}) with space indices Fourier transformed
into ${\bf q}$, but the energy integrals are performed.   An analytic expression for the surface Green's
function is available, which is $G^r_0({\bf p}) = - u/t$ with $u$ satisfying
\begin{equation}
 t + \bigl(E + i\eta - \epsilon({\bf p})\bigr) u + t u^2 = 0,\quad |u|< 1,
\end{equation}
and $\epsilon({\bf p}) = - 2 t \bigl( \cos(p_x a) + \cos(p_y a) \bigr)$ the electron dispersion relation
on a 2D square lattice. This is efficient numerically, and we were able to
compute large sizes of $L$ up to 255, which is sufficient for convergence to infinite sizes. 

\begin{figure} 
\includegraphics[width=\columnwidth]{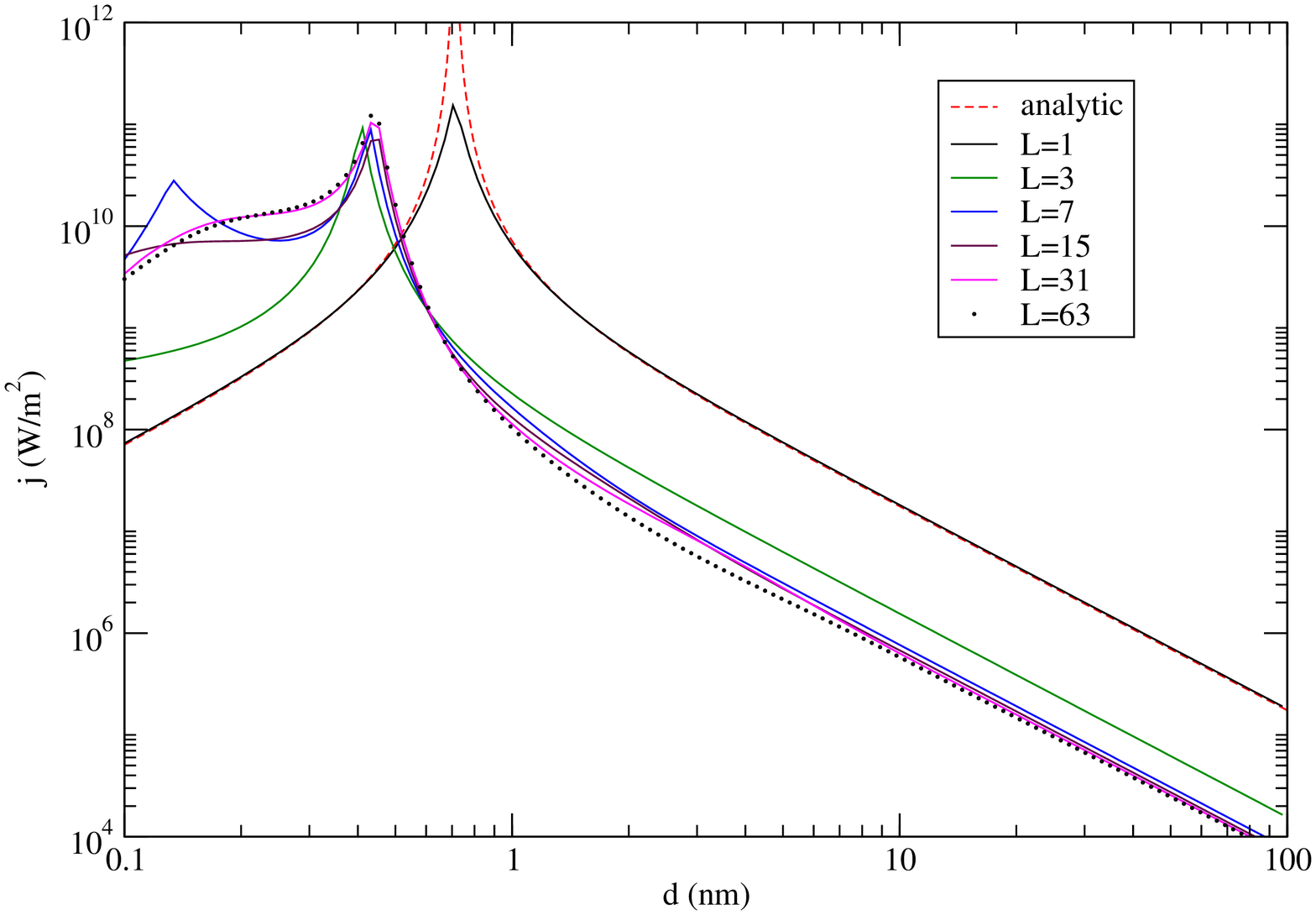}
\caption{The energy current density calculated by the Caroli formula as a function of tip to surface
distance for difference sizes of the surface blocks.  The dotted line indicated as `analytic' is from
the expression Eq.~(\ref{eq-analytic}), with $I_1/a^2$, $a=2.88\,$\AA.}
\label{fig-tip-surface}
\end{figure} 

In Fig.~\ref{fig-tip-surface}, we present numerical results for the model with the following parameters:
tight binding hopping parameter $t = 0.85\,$eV, lattice constant $a=2.88\,$\AA, a small
damping parameter in the solution for surface Green's function $\eta = 11\,$meV.   The temperatures
are $T_1 = 350\,$K and $T_2 = 300\,$K, and chemical potentials of both sides are set to 0.
We present the total current divided by $a^2$ so that thermal current density
$I_1/a^2$ can be compared with the black-body value (which
is $j_{BB} = 391.6\,$W/m$^2$ for our parameters) and the parallel plate cubic blocks in the next subsection.
As the sizes $L$ of the plane increase, the results quickly converged.  We have also calculated
up to $L=255$, but the results are nearly identical to that of $L=63$.  The last sizes in the figure
represent the limiting value of an infinite large surface.   We attribute the quick saturation to
the short screening length of the electron gas represented by the cubic lattice.
As for the general behavior of the distance dependences, it is very clear that current density
decays with distance as $1/d^2$, in agreement with analytic results for the two-dot model.   The short-distance
results should not be taken literally as at these distances electrons start to tunnel, and 
we expect the model to break down. 

\subsection{Cubic blocks}
The last example is cubic lattice blocks on both sides.  In this case, we have allowed the electrons
to hop to their nearest neighbors on the other side with a distance dependent hopping parameter 
$t \exp[-4 (d-a)/a]$, here $a = 2.88\,$\AA\ is the lattice constant.   Other parameters are
the same as the surface-tip problem.
We use the Meir-Wingreen formula to compute the energy current under $G_0 W_0$ 
approximation to the electron self-energy (omitting the Hartree term).
As can be seen, at short distances, the electron
tunneling induces huge thermal current without a large electric current (not shown).   
If $L=1$, when without a transverse direction,
the system is the same as the 1D chain point charges.   For $L=1$, 2, and 4, we compute using
real space Coulomb interaction formulation presented here.  For the curve labeled inf$\times$inf, it is
computed according to the method in Ref.~\onlinecite{zuquanPRB}.  We have used $80\times 80$ for the
$k$-point sampling, and used $2048$ points for the energy/time domain fast Fourier transform.
We like to see the effect of increasing $L$ and the
converged result when $L\to\infty$.     
To our surprise,
the near field enhancement is greatly diminished as $L$ becomes large.  When $L$ is practically
infinite, the magnitude is comparable to the standard PvH theory (as we should expect).   Thus,
small geometric form factor, less screening of electrons, is the reason why we see large enhancement
for the dot models. 

\begin{figure}
\includegraphics[width=\columnwidth]{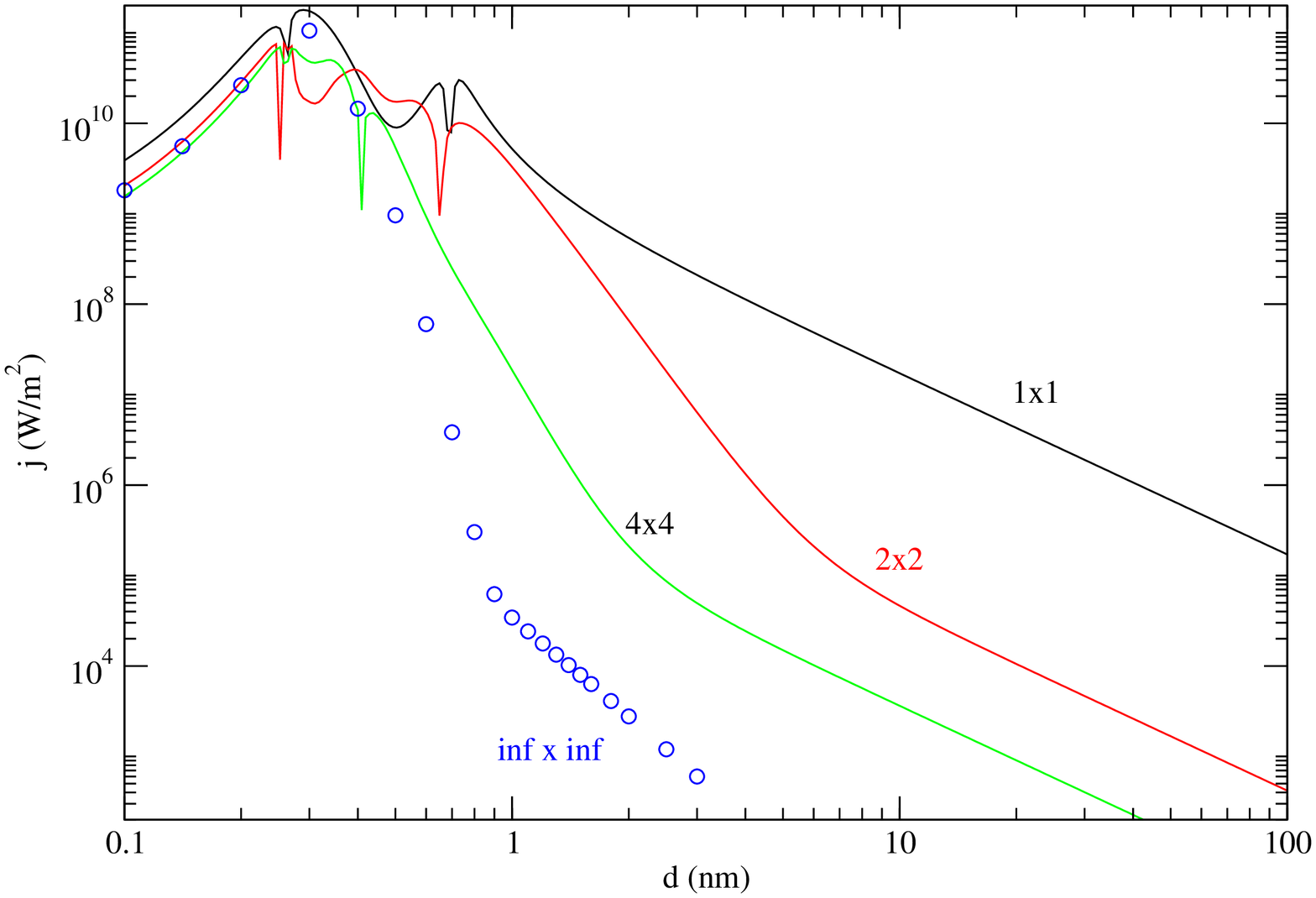}
\caption{The energy current density of cubic lattice blocks.  The circles indicated as `infxinf' are 
results from the wavevector space method.}
\label{cubic-lattice}
\end{figure}

\begin{figure}
\includegraphics[width=\columnwidth]{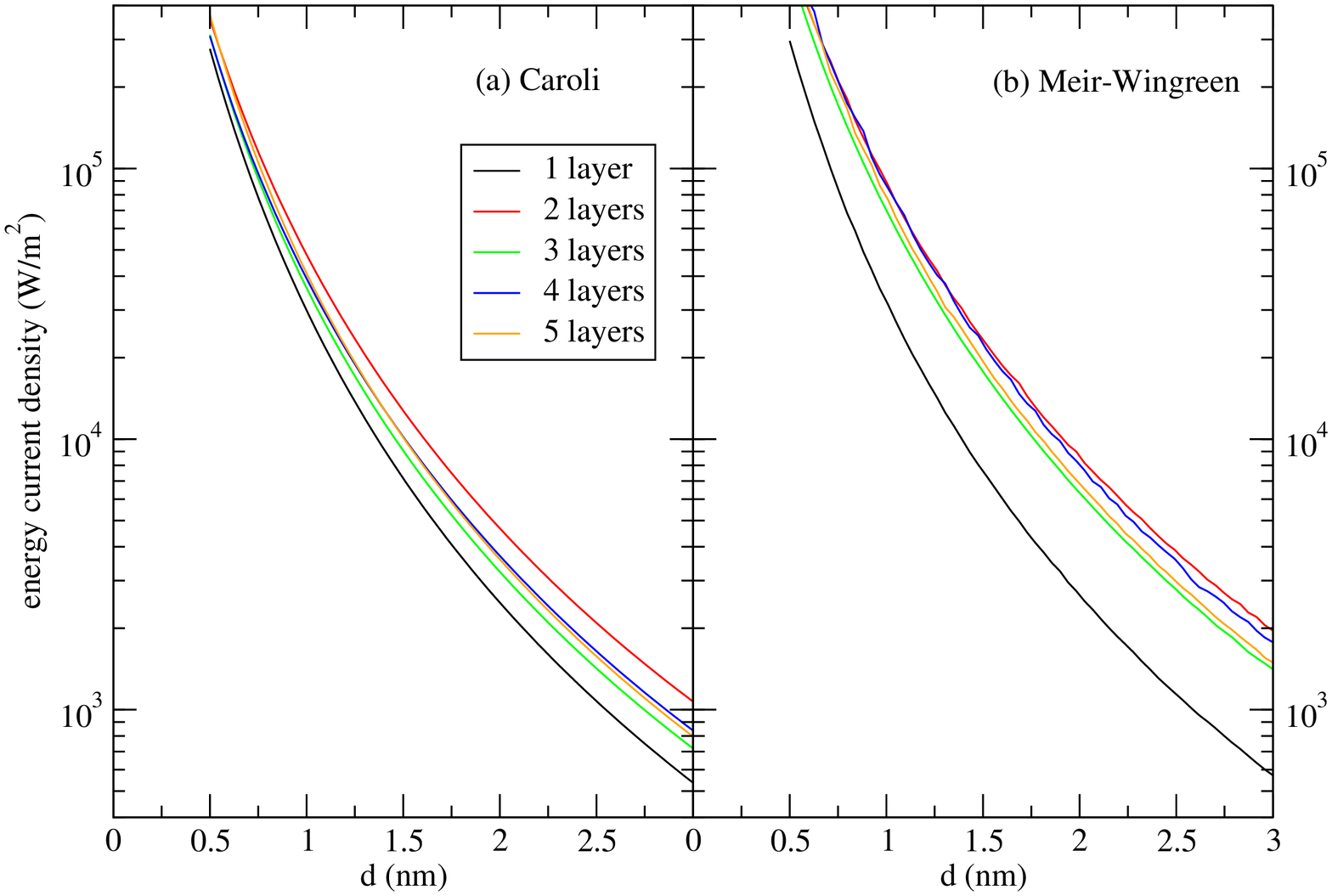}
\caption{Layer dependence of the energy current density of cubic lattice blocks without allowing the electrons to
tunnel.  The model parameters are, hopping $t=0.85\,$eV, lattice constant $a=2.88\,$\AA, 
temperatures $T_1 = 350\,$K and $T_2= 300\,$K.  
(a) From the Caroli formula, Eq.~(\ref{eq-landauer}), (b) from the Meir-Wingreen formula, Eq.~(\ref{eq-meir-wingreen}), using
$G_0 W_0$ for the electron Fock self-energy.  Legends are the same as for (a).
 The $k$-sample points are $36\times 36$ or more with 
6000 points or more for the Fourier transform.}
\label{cubiclayers}
\end{figure}

We have used the free electron models for the leads.  Is this well justified?  The answer is yes.
We can enlarge the center so that more layers are experiencing Coulomb interactions.  Figure~\ref{cubiclayers}
shows the numerical results of the layer dependences for the cubic block models without the
tunnel couplings.  Three to five layers are sufficient for a converged
result and it is not very much different (at most a factor of three) from a one-layer result.   
This is understandable from the
Thomas-Fermi screening.  The screening length in metal is rather short, usually of the order of
few lattice spacings.

For the one layer model, we found excellent agreement between the Caroli formula results and those
based on Meir-Wingreen.  However, for multi-layer cases, they differ.   This means that
the replacement  from $G^r$ to $G_0^r$ in applying Keldysh equation is not very good for the 
multi-layer case, an indication that local equilibrium assumption is likely less accurate.   Since
$G_0 W_0$ is not a norm conserving approximation, the current computed
from the left can differ from the right by 20 to 50 per cent.  What is plotted in Fig.~\ref{cubiclayers}(b) 
is the average, $\frac{1}{2}(I_1\! -\!I_2)/(aL)^2$.

\section{Summary and discussion}
We presented a simple and straightforward procedure to calculate ultra-near-field energy exchange
mediated by Coulomb interactions involving electrons.  The Caroli formula is valid when electrons
are not allowed to tunnel, while the Meir-Wingreen is needed when electron can tunnel.  In the
regime when electrons tunnel, the magnitude of energy transfer is comparable to typical heat
conduction.   An intriguing feature we found is that if electrons can be isolated (in the sense
that they can be modelled as quantum dots, with a strong system-bath coupling), much higher
near field enhancement is obtained.    To compare the quantum dot models with the surface
parallel plate geometry, we have divided the current by the area $a^2$ of a unit cell.   Clearly this normalization
into an energy current density is a bit arbitrary.   However, if we imagine packing a bunch of
1D chains into a 2D surface without introducing further Coulomb interaction, that number is
what we should get.  Unfortunately, Coulomb interaction does exist.  As a result, 3D lattice
with a 2D square lattice surface has a much reduced near field heat transfer.  

If our quantum dots do not represent a single electron, rather a group of electrons moving
in unison, then the effective charge will be $Q=-ne$ of $n$ electrons.  The energy transfer
will be proportional to $Q^4$,   so we expect a collective motion degree of freedom with net
charge larger than a single electron will have a much high energy transport.

\section*{Acknowledgements}
This work is supported by FRC grant R-144-000-343-112.

\appendix

\section{A derivation of the Caroli formula}
In this appendix we give a derivation of the Caroli formula in the spirit of Rytov fluctuational electrodynamics \cite{Rytov}.  
Since from the point of view of the NEGF Meir-Wingreen formula, the Caroli/Landauer formula cannot be an exact result, 
we would like to pinpoint where we have made an approximation.   The central idea of PvH is to generalize
Maxwell's equations into a stochastic form.  In the present context, it is the Poisson equation,
$-\epsilon_0 \nabla^2 \phi = \rho$.   Using the discrete formulation, we postulate
\begin{equation}
\label{eq-fluctuation-poisson}
v^{-1} \phi = \xi + \Pi^r \phi,
\end{equation}
here the scalar potential  $\phi$ and random noise $\xi$ are column vectors of size $N$, and the bare
Coulomb term $v$ and retarded self-energy $\Pi^r$ (in frequency domain) are $N \times N$ matrices.
The solution is readily obtained as $\phi = D^r \xi$ with $D^r$ satisfying a Dyson equation. 

We can give the retarded version of the Dyson equation in frequency domain, 
$D^r = v + v \Pi^r D^r$,  the following interpretation.
The bare Coulomb matrix $v$ maps charge into scalar potential, $\phi = v q$.  Moving the last term in the
Dyson equation to the left, we can write $\epsilon D^r = v$, and $\epsilon \equiv I - v \Pi^r$ is the
dielectric function matrix, here $I$ is the identity matrix.  
Thus, $D^r$ maps the external testing charge to the total screened potential,
$\phi = D^r q^{\rm ex}$.   The total charge in the system can be separated into two parts,
the external charge $q^{\rm ex}$ and induced charge $\delta q = \Pi^r \phi$.   We identify
this external charge as the random noise $\xi$ with $\langle \xi \rangle = 0$.   
The origin of the random charge is due to
the fact that the central region is not isolated.  The connections with the electron leads
result in random fluctuation of charges $\xi$ due to thermal agitations.   In order to have
a self-consistent description in the sense that $- (i/\hbar) \langle \phi(t) \phi(t')^T \rangle
 = (D^{>} \,+\, D^{<})/2$ reproduces the NEGF result of the Keldysh equation,
$D^{>,<} = D^r \Pi^{>,<} D^a$, we demand \cite{WangPRL07}
\begin{equation}
\frac{1}{i\hbar}\langle \xi(t) \xi(t')^T \rangle = \frac{ \Pi^{>}(t\!-\!t') + \Pi^{<}(t\!-\!t')}{2}
\equiv \bar{\Pi}(t\!-\!t').
\end{equation}
Here the averages $\langle \cdots\rangle$ are with respect to the random noises.

We consider a central region consisting of $N$ sites which can be separated into regions 1 and 2,
with $N_1 + N_2 = N$. Electrons are not allowed to tunnel between the two.  Then $\Pi^r$ is
decomposed as two block-diagonal matrices of $\Pi^r_1$ of $N_1 \times N_1$ and $\Pi^r_2$ of
$N_2 \times N_2$.

We consider the heat transfer by joule heating, ${\bf j} \cdot {\bf E}$, which after integration
by part, is $- \dot{\rho} \phi$.   Since Eq.~(\ref{eq-fluctuation-poisson}) is linear, we can consider
the effect of random noises of two sides separately.  Turning off $\xi_2$, the energy transfer
due to the fluctuation of charge of left side $\xi_1$ to the right side is
\begin{equation}
I_{1\to 2} = -\langle \dot{q}_2^T \phi_2\rangle,
\end{equation}
where $\phi_2 = D^{r}_{21} \xi_1$, and $q_2 = \Pi^r_2 \phi_2$.  These are time domain quantities,
for example, 
\begin{equation}
\phi_2(t) = \int D_{21}^r(t-t') \xi_1(t') dt'.
\end{equation}
We assume that the system is in steady state and $I_{1\to 2}$ is in fact independent of time.
Representing all the time domain quantities by their Fourier transforms in frequency domain, after
some lengthy but straightforward algebra, we find
\begin{equation}
\label{eq-I1to2}
I_{1\to 2} =  \int_{-\infty}^{+\infty} \frac{d\omega}{2\pi} \hbar \omega
{\rm Tr}\bigl(D^a_{12} \Pi^a_2 D^r_{21} \bar{\Pi}_1  \bigr).
\end{equation}
The last factor is due to noise correlation.  An important assumption going into the proof of the 
Caroli formula is to assume that the left and right sides are in respective equilibrium ---
we call this local equilibrium approximation.  Thus, we assume the fluctuation-dissipation
theorem, $i \bar{\Pi}_1 = (N_1 + 1/2) \Gamma_1$, which is valid for equilibrium systems, can be applied.  
Here $N_1$ is the Bose function at temperature $T_1$, and $\Gamma_1 = i (\Pi^r_1 -\Pi^a_1)$.

The energy pumped from 2 to 1 by $\xi_2$ can be obtained similarly by swapping the
index $1 \leftrightarrow 2$.   The overall heat current from left to right is given by the difference,
$I_1 = I_{1\to 2} - I_{2\to 1}$.  The expression can be simplified using the fact that
(1) $I_{1\to 2}$ and $I_{2\to 1}$ are real, so that we can take the Hermitian conjugate of
the factors inside the trace and add them, then divide by 2.  (2) We can perform cyclic permutation
under trace. (3)  Both $D^r$ and $\Pi^r$ are symmetric matrices, thus, e.g.,
$D^a = (D^r)^\dagger = (D^r)^{*}$.   With these manipulations, the expression can be
simplified to the standard Caroli form, Eq.~(\ref{eq-landauer}) and (\ref{eq-caroli}), 
in the main texts.

\section{Equivalence to Yu et al.}
The expression of Yu et al.\ uses the susceptibility $\chi$ which is related to the dielectric matrix by
$\epsilon^{-1} = I + v \chi$, or $\chi = \Pi^r \epsilon^{-1}$.   In terms of $\chi$, the
Dyson equation is $D^r = v + v \chi v$. We define the submatrices
$\epsilon_1 = I_1 - v_{11} \Pi_1^r$ and $\chi_1 = \Pi^r_1 \epsilon_1^{-1}$ of sizes $N_1 \times N_1$, 
and similarly for $\epsilon_2$ and $\chi_2$ of sizes $N_2 \times N_2$.  These quantities are the material
properties of system 1 and 2 in isolation.   The Dyson equation couples the two sides.  
We can write in the form $\epsilon D^r = v$, or in block matrix form
\begin{equation}
\left( \begin{array}{cc}
\epsilon_{1} &  -v_{12} \Pi^r_2 \\
-v_{21} \Pi^r_1   & \epsilon_{2}
\end{array}   
\right) 
\left( \begin{array}{cc}
D_{11} &  D_{12} \\
D_{21}   & D_{22}
\end{array}   
\right) = 
\left( \begin{array}{cc}
v_{11} &  v_{12} \\
v_{21}   & v_{22}
\end{array}   
\right).  
\end{equation}
Here $v_{11}$ and $v_{22}$ are the Coulomb interactions connecting the same side,
and $v_{12} = v_{21}^T$ connecting different sides.  Due to the diagonal nature of
$\Pi^r$, we do not need all the entries of $D^r$, only $D_{21}$.  Focusing on the first column, we 
obtain pair of equations,
\begin{eqnarray}
\epsilon_1 D_{11} - v_{12} \Pi^r_2 D_{21} = v_{11},\\
-v_{21} \Pi^r_1 D_{11} + \epsilon_2 D_{21} = v_{21}.
\end{eqnarray}
Eliminating $D_{11}$, we find
\begin{eqnarray}
\label{eq-D21}
D_{21} &=& \epsilon_2^{-1} \Delta_2^T v_{21} (\epsilon_1^{-1})^T, \\
\label{eq-Delta2}
{\rm with}\quad \Delta_2 &=& \bigl( I_2 - \chi_2 v_{21} \chi_{1} v_{12} \bigr)^{-1}.
\end{eqnarray}

In the last step, we find a relation between the imaginary part of $\Pi_1^r$  and
$\chi_1$, and similarly for system 2.  For notational simplicity, we drop the
subscripts 1 and 2 and script $r$  for the moment. From the relations $\epsilon = I - v \Pi$ 
and $\epsilon^{-1} = I + v \chi$,  multiplying $v^{-1}$ from left we obtain
\begin{equation}
v^{-1} \epsilon = v^{-1} -\Pi, \quad 
v^{-1} \epsilon^{-1} = v^{-1} + \chi.
\end{equation}
Taking the Hermitian conjugate of each, and then subtracting them, since $v$ is real symmetric, we get
\begin{eqnarray}
\label{eq-imaginary-Pi}
\Pi - \Pi^{\dagger} &=& -v^{-1} \epsilon + \epsilon^{\dagger} v^{-1}, \\
\chi - \chi^{\dagger} &=& v^{-1} \epsilon^{-1} - (\epsilon^{-1})^{\dagger} v^{-1}.
\end{eqnarray} 
Multiplying by $(\epsilon^\dagger)^{-1}$ from left and $\epsilon^{-1}$ from right to 
Eq.~(\ref{eq-imaginary-Pi}), and using the fact that both $\Pi$ and $\chi$ are symmetric
matrices, $\Pi - \Pi^\dagger = 2 i\, {\rm Im} \Pi$,  $\chi - \chi^\dagger = 2 i\, {\rm Im} \chi$, we find
\begin{equation}
\label{eq-Imchi}
 (\epsilon^\dagger)^{-1} {\rm Im} \Pi \,\epsilon^{-1} = {\rm Im} \chi.
\end{equation} 
Taking the transpose of the equation, and $\chi = \chi^T$, we also have
$ (\epsilon^{-1})^{T} {\rm Im} \Pi \,(\epsilon^{-1})^{*} = {\rm Im} \chi$.
Finally, putting the expressions together, Eq.~(\ref{eq-D21}) and (\ref{eq-Imchi}),
the transmission function is,
\begin{eqnarray}
T(\omega) &=& T(\omega)^{*}   \nonumber \\
&=& 4\, {\rm Tr} \bigl(D_{21} {\rm Im} \Pi_1 ( D_{21})^{\dagger} {\rm Im} \Pi_2   \bigr)^{*}  \nonumber \\
& = & 4\, {\rm Tr} \bigr( \Delta_2^\dagger v_{21} {\rm Im} \chi_1 v_{12} \Delta_2 {\rm Im} \chi_2   \bigr)
\end{eqnarray} 
We have used the fact that the transmission function is real, and did a cyclic permutation of a
term under trace.  This is the form given by Yu et al.~\cite{Yu2017}, their Eq.~(8).

\bibliography{RHTNEGF}

\end{document}